\documentclass[12pt]{elsarticle}
\usepackage[utf8]{inputenc}

\usepackage{graphicx}
\usepackage{epsfig}
\usepackage{amssymb}
\usepackage[section]{placeins}
\usepackage{latexsym}
\usepackage{amsmath}
\usepackage{subfiles}
\usepackage{siunitx}
\usepackage{xcolor}
\usepackage{hyperref}

\newcommand{\rfig}[1]{Fig.~\ref{#1}}    
\newcommand{\rsec}[1]{Sec.~\ref{#1}}    

\begin{document}
\begin{frontmatter} 

\title{On the synergy between easier plasma operation and affordable coil-set requirements enabled by Negative Triangularity in the prospective ARC fusion reactor}
\author[label1]{N. de Boucaud}
\author[label2]{T. Golfinopoulos, A. Marinoni}
\date{December 2022}

\affiliation[label1]{organization={General Atomics},
            addressline={3550 General Atomics ct}, 
            city={San Diego},
            postcode={92121}, 
            state={CA},
            country={USA}}
\affiliation[label2]{organization={Massachusetts Institute of Technology, Plasma Science and Fusion Center},
            addressline={77 Massachusetts Avenue, NW17}, 
            city={Cambridge},
            postcode={02139}, 
            state={MA},
            country={USA}}
\maketitle            
\begin{abstract}
A numerical workflow is developed to explore the viability of running multiple plasma configurations in the ARC fusion pilot plant. Suitable cost functions for various poloidal field coil sets are evaluated based on currents required in the coils, induced stresses, and flexibility in plasma configurations. It is shown, for the first time, that a given set of poloidal field coils can sustain equilibria with both signs of triangularity and that equilibria at Negative Triangularity have comparable coil requirements to those at Positive Triangularity, despite ARC's Positive Triangularity design equilibrium. These results contribute to demonstrate that the Negative Triangularity Tokamak is a promising candidate for the commercialization of fusion power.  
\end{abstract}






\tableofcontents

\end{frontmatter}

\section{Introduction}\label{SEC:intro}

The Tokamak is currently considered the leading candidate device for an economically viable nuclear fusion reactor. 
Conceptualized in the 1950s by Soviet physicists I.~Tamm and
A.~Sakharov~\cite{Shafranov:JRAS2001}, tokamaks confine hot plasmas in a toroidal vacuum chamber by using a helical magnetic field. In equilibrium, the plasma is organized in a number of nested isobaric surfaces. 
Many of the early tokamaks, starting with the first-ever-built device
\textit{T-1}, employed a near-circular poloidal cross section, which was the
natural choice one would make when folding the cylindrical plasma employed in linear devices into a torus. Since the late 1960s-early 1970s, however, the fusion community started deforming the circular cross section with the goal of improving performance by sustaining more current for the same value of the kink safety factor~\cite{Solovev:PFCNFR1969,Laval:PFCNFR1971,Coppi:PF1972,Artsimovich:JETP1972}. Subsequent theoretical work in the realm of Magneto-Hydro-Dynamic stability demonstrated that ballooning modes, a localized pressure-driven instability, can be stabilized at higher pressure than that at which they first become unstable~\cite{Mercier:PPCNFR1978,Lortz:PL1978,Coppi:PRL1980}: a phenomenon known as second stability. It was also found that, by adopting \textit{dee-shaped} cross sections, also commonly referred to as elongated shape with \textit{positive triangularity} (PosT), the two regions of stability are connected thereby allowing one to operate at higher pressure gradients provided that turbulent transport is somehow quenched. In the H-mode regime, discovered in the early 1980s~\cite{Wagner:PRL1982}, edge turbulence is stabilized by flow-shear~\cite{Biglari:PF1990,Burrell:PoP1997}, which permits the creation of an edge insulating layer, dubbed pedestal, with steep pressure profiles near the plasma periphery. The plasma core, although still turbulent, thereby achieves the high pressure conditions necessary for self-sustained nuclear fusion.  
The ability of the dee shape to sustain a
stable plasma to higher pressure was reported by numerous devices
worldwide~\cite{Troyon:1988,Neyatani:PPCF1996,Taylor:PPCF1997,Hender:2007} thus providing a convincing argument that the H-mode regime should be pursued in future fusion reactors. Plasmas at positive triangularity were also attractive on technological grounds because, according to the \textit{Princeton D-coil} design~\cite{File:IEEE1971}, mechanical stresses in the plane of the
toroidal field coils are minimized when the coil is shaped as a \textit{dee}.
When considering stresses out of the plane of the coil, however, the \textit{Princeton D-coil} shape is no longer an optimal solution~\cite{Gray:ORAU1974}.
\par
In H-mode plasmas the height of the edge pedestal correlates with the confinement level which, in turn, is inversely proportional to the capital cost of a power plant for a given power output. Since the pedestal height generally increases at increasing triangularity thanks to the stabilization of peeling-ballooning modes~\cite{Wilson:PoP2002}, part of the fusion community is actively working towards reactor solutions such that the pedestal height is maximized~\cite{Snyder:NF2019}. However, the strategy of maximizing the pedestal height, and the H-mode in general, feature intrinsic operational challenges. First, pedestals cause the triggering of edge localized modes (ELMs), a bursting instability that must be actively suppressed or, at the very least, mitigated~\cite{Lang:NF2013} to avoid the deterioration of plasma facing components (PFCs) \cite{Loarte:NF2014}. Second, the power flow crossing the plasma edge must remain above the L$\rightarrow$H power threshold for the high confinement state to persist. Even in the absence of ELMs, such power flow in future reactors will have to be dissipated near the divertor in order to shield PFCs~\cite{Soukhanovskii:PPCF2017} in a configuration called \textit{detachment}~\cite{Krasheninnikov:JPP2017}. However, pedestals are extremely sensitive to the position of the detachment front and on the concentration of impurities which are advected inside the plasma by the main ion density gradient. Finally, the impurity retention caused by pedestals excessive amounts of impurities lowers fusion performance due to excessive dilution of the main ion species. \par
All these issues have recently made part of the fusion community question whether the H-mode regime is a viable candidate for operation in future fusion reactors. As a result, alternative magnetic configurations are being explored, or revisited, to look for optimal solutions using the more advanced numerical tools that are now available.
Cross sectional shapes characterized by a \textit{reversed-dee}
configuration, also known as \textit{negative triangularity} (NegT), were examined and quickly dismissed on grounds of poor MHD stability. Nevertheless, recent experiments on the TCV~\cite{Coda:APS2019} and DIII-D~\cite{Austin:PRL2019,Marinoni:PoP2019, Marinoni:RMPP2021,Marinoni:NF2021} tokamaks showed promising features that would ease operations in future reactors. Indeed, while the stability of ballooning modes in NegT plasmas is such that second stability cannot be achieved, thereby preventing H-mode access~\cite{Saarelma:PPCF2021}, the intensity of turbulent fluctuations in NegT configurations is decreased to a level such that H-mode grade confinement and pressure levels are maintained even in the absence of an edge pedestal. These results motivate this paper, which explores the viability of equilibria at NegT in the ARC~\cite{Sorbom:SORBOM2015378} design and associated impacts on the poloidal field coils that will be necessary to obtain such configurations.\par
The paper is organized as follows. Sec. \ref{SEC:wf} describes the details of the numerical workflow used, Sec. \ref{SEC:res} illustrates the results obtained, while Sec. \ref{SEC:conc} summarizes the paper.
\section{Numerical workflow}\label{SEC:wf}
\begin{figure}[!htb]
    \centering
    \includegraphics[width=12cm]{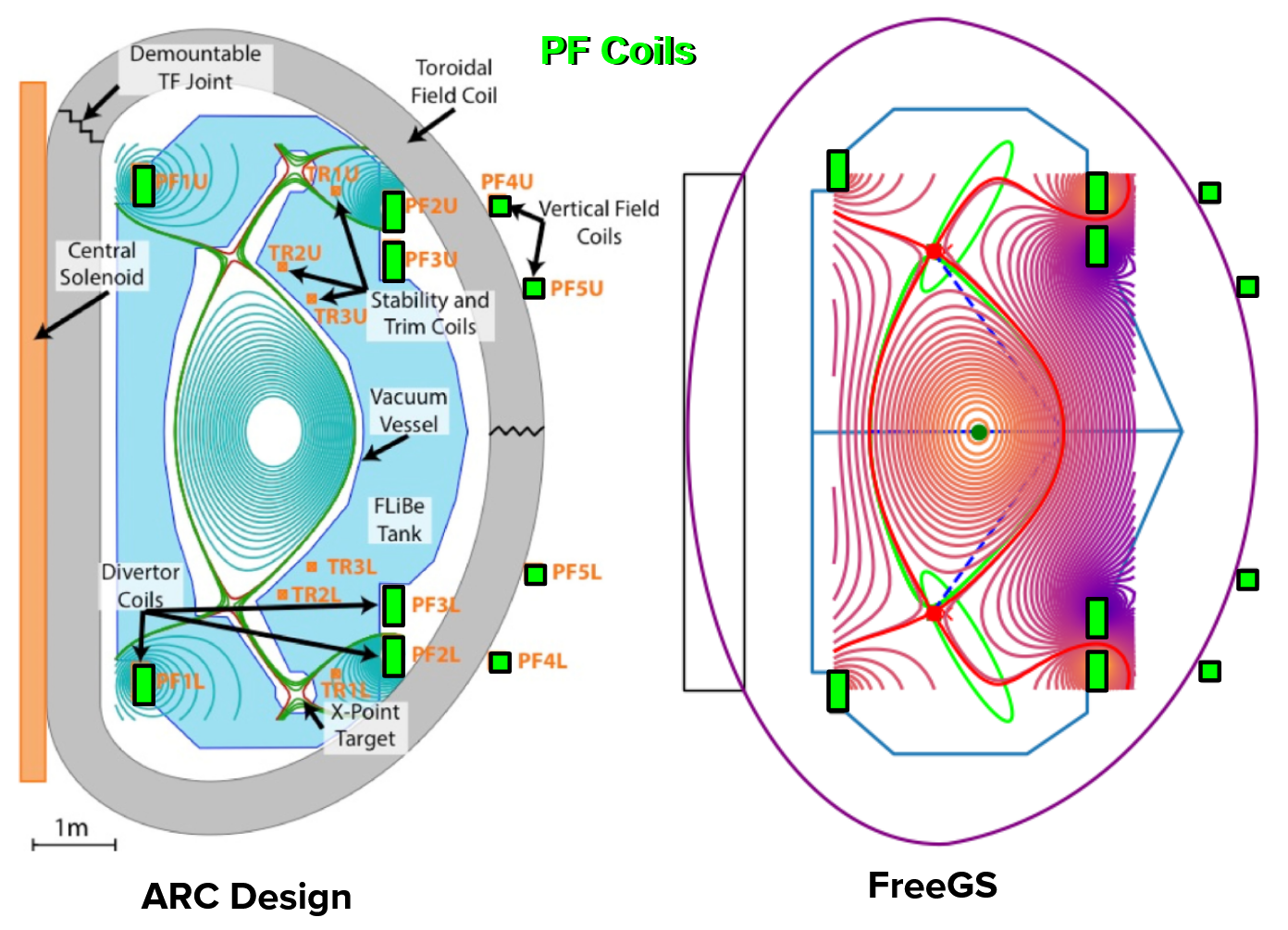}
    \caption{ARC Baseline Design Compared with FreeGS Reconstruction}
    \label{fig:ARCcomp}
\end{figure}

In order to investigate the operational space of the ARC design, the time-independent Grad-Shafranov equilbrium solver FreeGS~\cite{Dudson:DUDSON2022} was employed. 
 The choice to adopt FreeGS was dictated by the fact that it is an open source code written in Python, which is therefore easy to interface with wrappers, and that it is able to generate solutions in a few seconds on a modern laptop. FreeGS was installed and properly tested using given example code. To generate plasma equilibria, wrapper functions were created that specify coil locations, constraints on the position of the separatrix, various plasma parameters, as well as on the plasma pressure and current spatial profiles. The accuracy of the reconstructed equilibria was validated against a few tens of DIII-D discharges whose equilibria were generated using the EFIT code~\cite{Lao:NF1985}. Once convergence was verified, wrapper scripts were built to interface with FreeGS. These scripts allow inputs to be sent to FreeGS in a batch and the output equilibria to be analyzed.
The baseline coil-set design of ARC (including its central solenoid) was built in FreeGS and its corresponding equilibrium was also validated using the created Python workflow by reproducing it from its published design parameters. Additionally, regions were created in the reactor to represent the FLiBe (molten salt) tank, divertors and toroidal field coil cross section. These regions were used in evaluating the cost of coil locations, as explained in \rsec{sec:cost} but are not taken into account by the solver. We note that, while higher order corrections caused by such components are beyond the scope of our study, the final design of a reactor will have to evaluate such effects. 
Shown in Figure \ref{fig:ARCcomp} is the baseline design of ARC compared side-by-side with the FreeGS reconstruction of it. The poloidal field coil currents from the baseline ARC reconstruction are shown in Table \ref{tab:ARCcomp}. Note that PF coils labeled "U" are located above the midplane of the machine while those labeled "L" are below it.
\begin{table}[h]
\caption{Comparison between the ARC baseline design and the FreeGS solution in terms of currents in the poloidal field coils in units of Ampere.}
\label{tab:ARCcomp}
\centering
\begin{tabular}{cccc}
PF Coil & ARC Baseline Design & FreeGS Solution & \% Error \\
\hline
PF1U & 3.9E+06 & 3.8E+06 & 2.6\%\\
PF2U & 5.2E+06 & 5.2E+06 & 0.0\% \\
PF3U & -4.4E+06 & -4.5E+06 & 2.3\% \\
PF4U & -1.4E+06 & -1.0E+06 & 29\% \\
PF5U & -3.5E+06 & -2.7E+06 & 23\% \\
PF1L & 3.9E+06 & 3.8E+06 & 2.6\% \\
PF2L & 5.2E+06 & 5.2E+06 & 0.0\% \\
PF3L & -4.4E+06 & -4.5E+06 & 2.3\% \\
PF4L & -1.4E+06 & -9.8E+05 & 30\% \\
PF5L & -3.5E+06 & -2.7E+06 & 23\% \\
\end{tabular}
\end{table}
\par
\begin{figure}[!htb]
    \centering
    \includegraphics[width=10cm]{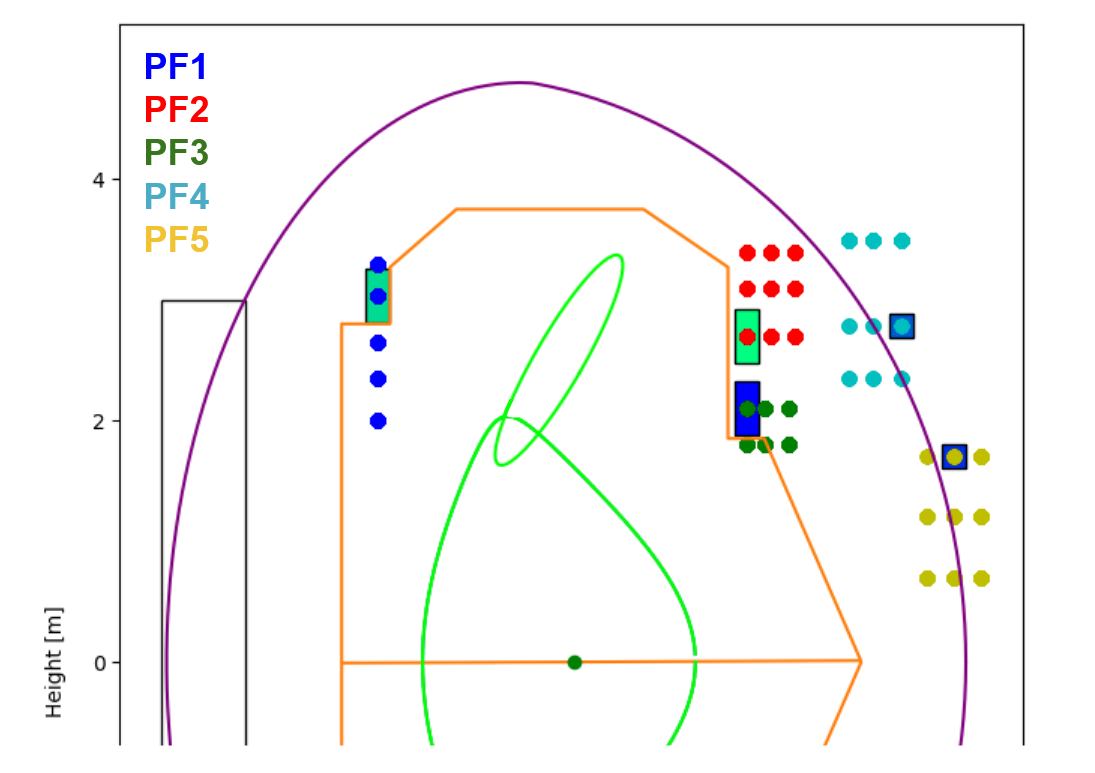}
    \caption{Search space for the location of poloidal field coils. Background filled positions represent the baseline ARC design.}
    \label{fig:coilvaries}
\end{figure}

With the ARC coil layout and equilibrium separatrix shape as a starting point, a search space was constructed to investigate the effects of varying the coil layout and separatrix shape. The shape of the separatrix was varied by altering both elongation and triangularity, with the latter allowed to assume negative values. In order to keep the number of permutations to a manageable level, the location of the coils is kept up-down symmetric, meaning that any given displacement of any of the coils above the midplane is exactly replicated below the the midplane. The position of each poloidal field coil was varied within the search space shown in Figure~\ref{fig:coilvaries}. This plot displays the FreeGS model of ARC wherein each color-coded point represents a different coil location for poloidal field coil 1 (PF1) through poloidal field coil 5 (PF5). Coil locations are mirrored about the midplane of the reactor.

\begin{figure}[!htb]
    \centering
    \includegraphics[width=10cm]{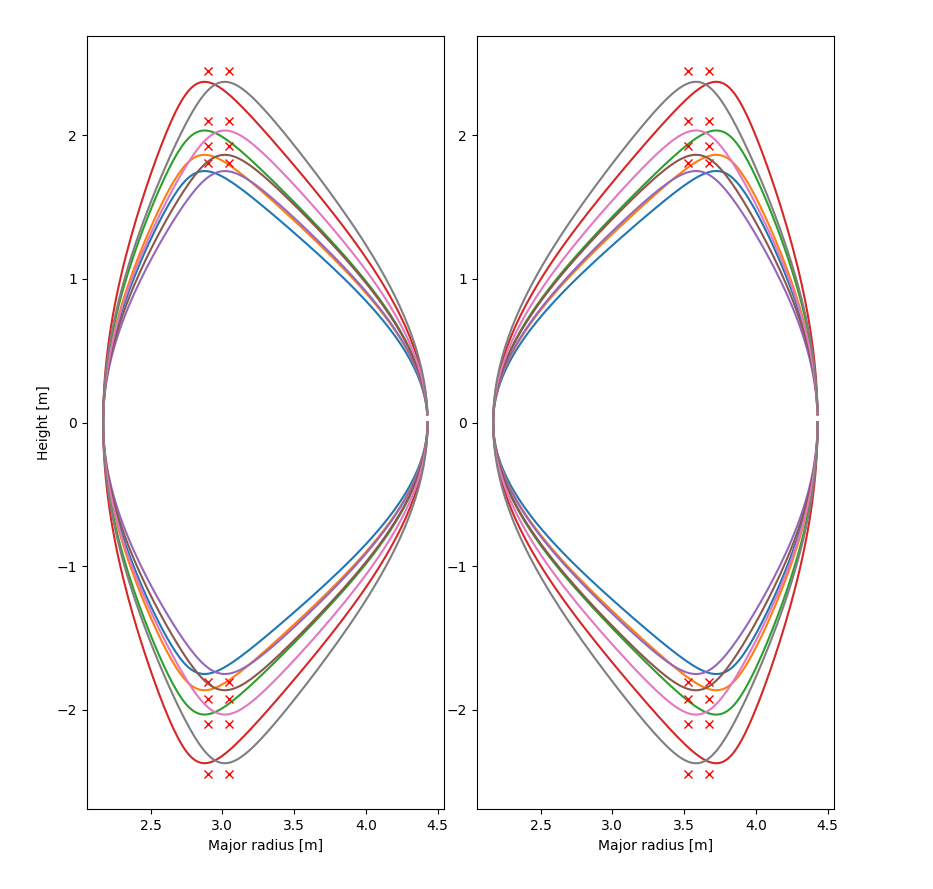}
    \caption{Visual representation of the various shapes of the separatrix that have been considered.}
    \label{fig:shapes}
\end{figure}

For each coil set tested, 16 separatrix shapes were simulated, covering a square grid in elongation $\kappa\in [1.55,1.65,1.8,2.1]$ and triangularity $\delta\in [-0.375,-0.25,0.25,0.375]$, thereby evaluating both positive and negative triangularity solutions for each value of elongation. All other plasma parameters remained the same throughout the simulations, including major radius: $3.3m$, minor radius: $1.13m$ and squareness: $-0.25$ The search space was chosen in such a way as to cover a broad range of shapes that are produced in current devices, in an effort to demonstrate that a given coil set can sustain both positive and negative triangularity configurations. More extreme values of triangularity as well as a more refined grid space in elongation are beyond the scope of this work and thus deferred to a future optimization.
A visual representation of the resulting plasma shape configurations is displayed in \rfig{fig:shapes}.
%

The number of combinations of the location of the poloidal field coils amounts to 21,870 which, including 16 shape variations for each coil configuration, generates a database of 349,920 cases. 
The total simulation time was about about 1 thousand CPU-hours, or less than four days on a modern CPU. The goodness of any given solution can be evaluated under a number of different metrics, such as the amount of current required in each coil or the sensitivity of the resulting equilibrium to the coil location; \rsec{sec:cost} details the procedure adopted in this work. 

\newpage
\subsection{Definition of the cost function}\label{sec:cost}

Each case was evaluated based on a cost function taking into account the current in the PF coils, the forces on the PF coils and the fit of the solution separatrix shape to the desired shape. Simultaneously, another cost function evaluated the coil layout in its performance over all 12 shapes by punishing coil sets with low flexibility of plasma shape, high average cost or PF coils in undesirable locations.

Evaluation of each case was done using equation (1) below.

\begin{equation}
    C = C_I + C_{hoop} + C_{Fz} + C_{match}
\end{equation}

Where $C_I$ is the cost of current in the poloidal field coils, $C_{hoop}$ is the cost of the hoop stresses acting on the coils, $C_{Fz}$ is the cost of the forces in the $z$ direction acting on the coils and $C_{match}$ is a cost measure of how well the coil set is able to produce a particular shape. Each of these costs is computed via an exponential function, normalized between values of 0 and 1 within a manually-tuned range of input parameter values. These exponential functions are shown in equation set (2) below.

\begin{equation}
\label{EQ:costs}
\begin{aligned}
C_I &= c_0\left(\frac{1}{c_0}\right)^{\frac{I}{I_0}}\\
C_{hoop} &= c_1\left(\frac{1}{c_1}\right)^{\frac{\sigma}{\sigma_{yield}}}\\
C_{Fz} &= c_2\left(\frac{1}{c_2}\right)^{\frac{F_z}{F_{z0}}}\\
C_{match} &= c_3\left(\frac{1}{c_3}\right)^{\frac{d}{d_0}}\\
\end{aligned}
\end{equation}

Where $c_0$, $c_1$, $c_2$, $c_3$, $I_0$, $F_{z0}$ and $d_0$ are manually tuned parameters which were found to give a desirable cost function behavior for evaluating coil sets and $\sigma_{yield}$ is the yield strength of REBCO coils~\cite{Barth:Barth_2015, Radcliff:2018_Fall_Radcliff_fsu_0071N_14949} (REBCO is the coil material used in the baseline ARC design \cite{ARC1}). $I$ is the absolute value of the current in a given coil (the maximum cost found is used to compute the cost in equation (1)). Similarly, $\sigma$ is the hoop stress computed from the radial force on the coil, found by FreeGS and $F_z$ is the force in the $z$ direction found by FreeGS (likewise, the maximum costs found are used to compute the cost in equation (1)). Lastly, $d$ is the root mean sum of the deviation of points in the computed separatrix from points in the desired separatrix shape.

From there, every coil set is evaluated based on another cost function which combines the results of each of the 12 separatrix shapes tested. This cost function has the form given by equation (3) below.

\begin{equation}
    C_{coilset} = C_{flex} + C_{avg} + C_{loc}
\end{equation}

Where $C_{flex}$ is the coil set's flexibility computed by equation (4) below.

\begin{equation}
    C_{flex} = 1 - \frac{n}{n_0}
\end{equation}

Where $n_0$ is the number of separatrix cases tested per coil set and $n$ is the number that were successful. A successful case is defined as one with a cost (equation (1)) less than a manually tuned maximum cost $C_{max}$ (a value of 20 was used for the results presented). This relationship is shown in equation (5) below.

\begin{equation}
    C_I + C_{hoop} + C_{Fz} + C_{match} = C < C_{max}
\end{equation}

$C_{avg}$ is the mean cost of every successful case. Cases with cost above the maximum negatively impact the cost of the coil set by raising the $C_{flex}$ metric but their affect isn't doubly accounted for since they aren't included in $C_{avg}$. $C_{loc}$ is the average cost of the coil locations. Computing the cost of an individual coil location was done by assigning cost values to each region previously defined in the ARC reactor.


\section{Results}\label{SEC:res}
\begin{figure}[!htb]
    \centering
    \includegraphics[width=14cm]{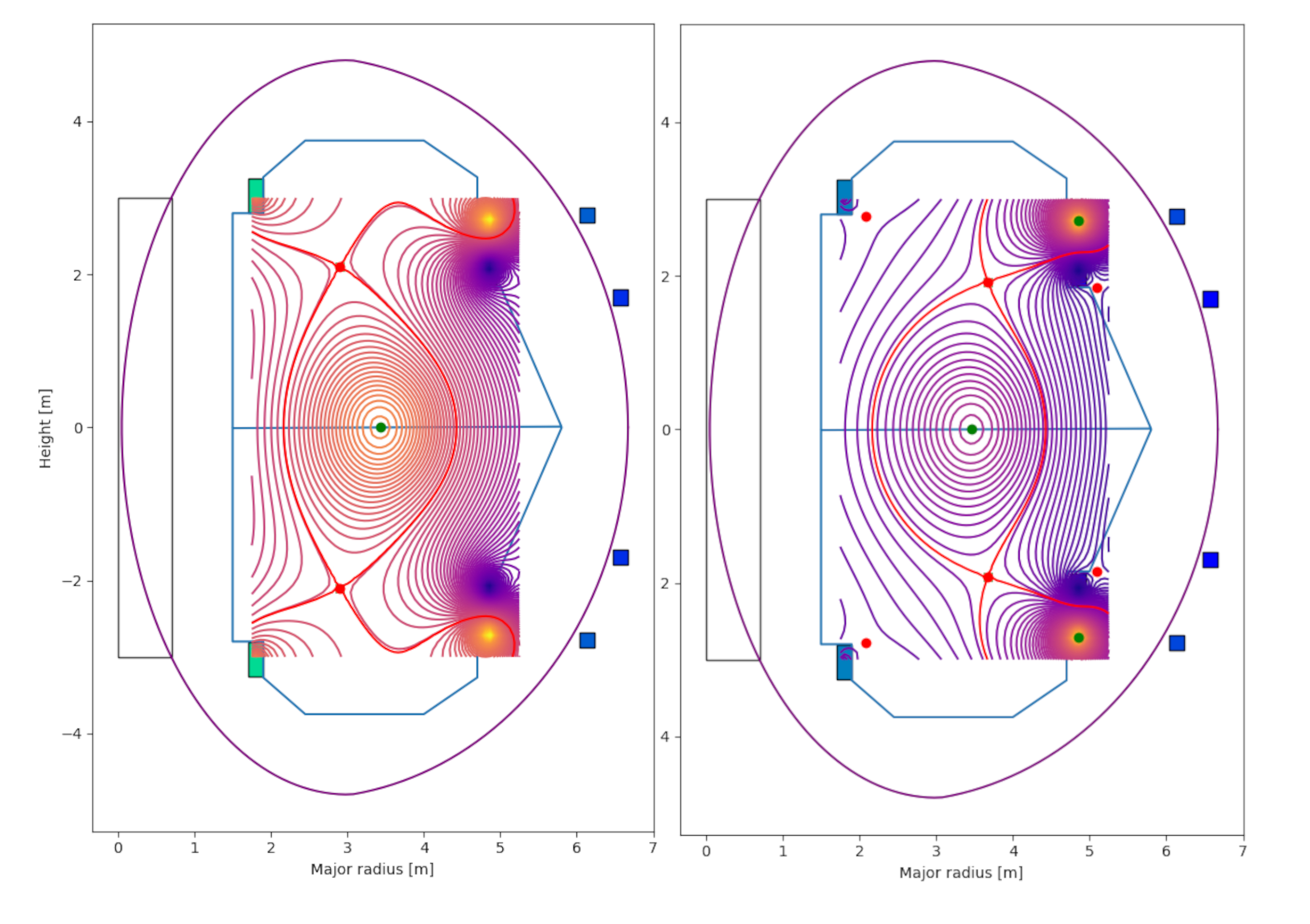}
    \caption{ARC baseline coil set with a PosT equilibrium (left) and a NegT equilibrium (right).}
    \label{fig:post&negt}
\end{figure}

As a result of the study, it was found that the ARC baseline coil set, and many variations of it, are able to produce plasma equilibria with both positive and negative triangularity. Plotted in \rfig{fig:post&negt} is the baseline ARC coil set showing this result. In order to reduce the amount of cases in the analysis with a poor separatrix shape result (i.e. one that failed to match the desired shape well), a limit was set on the root mean sum of the deviation of the separatrix from the desired shape. This limit was set to $0.1525$m, or 1 standard deviation above the mean. From this adjusted data set, PosT cases had an average total current in the PF coils of $35.5$ MA while NegT cases had $43.8$ MA, about $23$\% higher. PosT cases had an average z-directed force (averaged over all coils of each case) of $125.4$ MN while NegT cases had $166.4$ MN, about $33$\% higher. Likewise, PosT cases had an average hoop stress of $232.5$ MPa and NegT had $251.2$ MPa, about $8$\% higher.

This being the first use of the created costing workflow, little was known about how the cost functions would behave throughout a large number of cases. There is still much room to improve the evaluation metrics in order to approach a cost function that could find an optimal coil set and plasma shape for a fusion pilot plant at negative triangularity. This costing workflow in its current state was useful for evaluating outputs from FreeGS on a case-by-case basis in order to remove cases from the data set and for understanding of the main drivers of cost in the workflow.
%
%
\section{Conclusions}\label{SEC:conc}

A python based workflow was developed around the FreeGS solver to evaluate whether a given poloidal field coil set can be used to generate a variety of equilibria in the ARC pilot plant. The workflow was validated against discharges from the DIII-D database. The properties of the poloidal field coils were taken from the ARC base case study~\cite{Sorbom:SORBOM2015378, Kuang:KUANG2018221}, and their location varied while keeping up-down symmetry and suitable distance from the plasma to allow room for the shielding blanket. 
It is shown that the same coil-set allows to operate a variety of equilibria at varying elongation and triangularity, both positive and negative. Equilibria with a negative triangularity cross sectional shape are shown to have a comparable engineering requirements on average to those with positive triangularity shapes.

Negative triangularity configurations have been shown to have favorable properties both in terms of core confinement and power exhaust, which makes them viable candidates for operation in future fusion reactors~\cite{Coda:APS2019, Austin:PRL2019,Marinoni:PoP2019, Marinoni:RMPP2021,Marinoni:NF2021, Saarelma:PPCF2021}. This work demonstrates that negative triangularity plasmas can be attained on the ARC baseline coil set and too the potential for a Negative Triangularity optimized coil set to ease hardware requirements on the poloidal field coils.

\section*{Acknowledgements}
This work was supported in part by the U.S. Department of Energy under the Science Undergraduate Laboratory Internship (SULI) program and contracts  DE-SC0016154, DE-SC0014264. The authors wish to thank Dr. A.J. Creely for useful discussions as well as Dr. B. Dudson for the open source FreeGS code.

\bibliography{refs}
\bibliographystyle{elsarticle-num-names}
\end{document}